\documentstyle[prl,aps,graphicx]{revtex}
\begin{document}
\draft

\def\overlay#1#2{\setbox0=\hbox{#1}\setbox1=\hbox to \wd0{\hss #2\hss}#1%
\hskip
-2\wd0\copy1}
\twocolumn[
\hsize\textwidth\columnwidth\hsize\csname@twocolumnfalse\endcsname

\title{Quantum catastrophe of slow light}
\author{Ulf Leonhardt}
\address{School of Physics and Astronomy, University of St Andrews,
North Haugh, St Andrews, Fife, KY16 9SS, Scotland} \maketitle
\vskip2pc] \narrowtext

%%%%
{\bf Catastrophes are at the heart of many fascinating optical
phenomena$^{1}$. The rainbow$^{1}$, for example, is a ray
catastrophe where light rays become infinitely intense. The wave
nature of light resolves the infinities of ray catastrophes$^{1}$
while drawing delicate interference patterns such as the
supernumerary arcs of the rainbow$^{1}$. Black holes$^{2}$ cause
wave singularities. Waves oscillate with infinitely small wave
lengths at the event horizon$^{2}$ where time stands still. The
quantum nature of light evades this higher level of catastrophic
behaviour while producing a quantum phenomenon known as Hawking
radiation$^{3}$. As this letter describes, light brought to a
standstill in laboratory experiments$^{4,5,6}$ can suffer a
similar wave singularity caused by a parabolic profile of the
group velocity$^{7}$. In turn, the quantum vacuum is forced to
create photon pairs with a characteristic spectrum. The idea may
initiate a theory of quantum catastrophes, in addition to
classical catastrophe theory$^{8,9}$, and the proposed experiment
may lead to the first direct observation of a phenomenon related
to Hawking radiation$^{3}$.}

Optical media govern the propagation of light. Media are
transparent substances such as glass or water, but empty yet
curved space is a medium as well$^{10}$. One can manipulate
certain material media to give them extraordinary optical
properties. Inside such media light may propagate with a
negative$^{11}$ or very low$^{12}$ group velocity$^{7}$ or light
may be completely frozen$^{4,5,6}$. In a medium with
Electromagnetically-Induced Transparency$^{13}$ (EIT) an external
control beam dictates the group velocity $v_g$ of a second and
weaker probe beam to slow down the probe light$^{4,5,6,12}$. Once
the first beam has gained control, the group velocity of the
second one is essentially proportional to the control intensity
$I_c$, even in the limit when $I_c$ vanishes$^{14}$.

Imagine that the control beam illuminates the EIT medium from
above. Initially, the control intensity is uniform, but then the
control light develops a dark stripe. The stripe shall continue
down through the medium as an interface ${\cal Z}$ of zero
intensity $I_c$ where, consequently, the group velocity of any
potential probe light vanishes. In the following we show
theoretically that the interface ${\cal Z}$ forms the optical
analog of an event horizon. The formation of the horizon should
trigger a quantum catastrophe that results in a continuous
emission of slow-light quanta out of the vacuum. First we analyze
classical waves of slow light and then we turn to the quantum
theory.

\begin{figure}[htbp]
\begin{center}
\includegraphics[width=20.5pc]{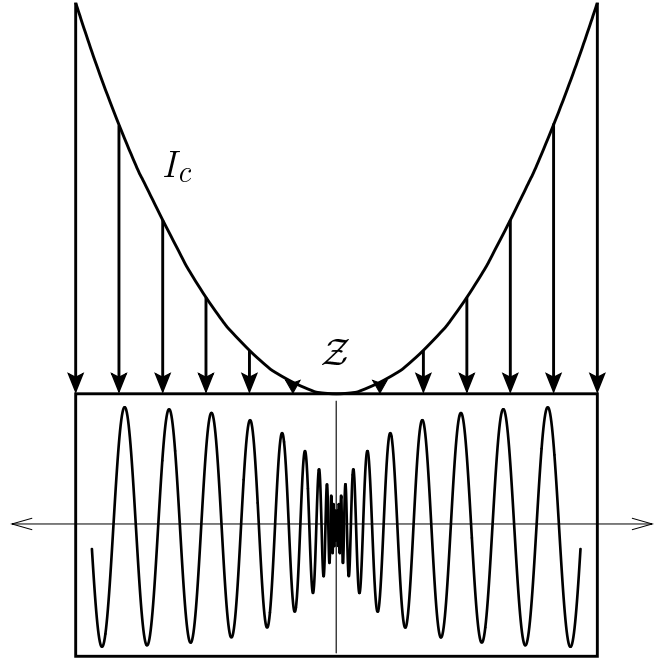}
\vspace*{2mm} \caption{Schematic diagram of the proposed
experiment. A beam of control light with intensity $I_c$ generates
Electromagnetically-Induced Transparency$^{13}$ in a medium,
strongly modifying its optical properties for a second field of
slow light. When an initially uniform control intensity is turned
into the parabolic profile shown in the figure, the slow-light
field suffers a quantum catastrophe. To slow-light waves, the
interface ${\cal Z}$ of zero control intensity cuts space into two
disconnected regions and creates a logarithmic phase singularity,
in analogy to the effect$^{18}$ of an event horizon$^{2}$. The
quantum vacuum of slow light cannot occupy such catastrophic
waves. In turn, pairs of slow-light quanta, propagating in
opposite directions away from ${\cal Z}$, are emitted with a
characteristic spectrum. The waves shown below the intensity
profile refer to the emitted light with the modes $w_R$ and $w_L$
of Eq.\ (\ref{set}). } \label{figure1}
\end{center}
\end{figure}

Assume that the interface ${\cal Z}$ of zero control intensity
$I_c$ is sufficiently flat such that the optical properties
generated do not vary much in the spatial directions parallel to
${\cal Z}$. Consider a line $z$ orthogonal to ${\cal Z}$. Over a
small fraction of a characteristic length $a$, the group-velocity
profile of the slow probe light is parabolic,
%%%%%%
\begin{equation}
\label{v} v_g \sim c\,\frac{z^2}{a^2}\,,
\end{equation}
%%%%%%
\newpage
\noindent because $I_c$ increases quadratically in the vicinity
of a zero. As usual, $c$ denotes the speed of light in vacuum.
For simplicity we concentrate on slow-light waves $\varphi(t,z)$
that propagate in $z$ direction only and we ignore their
polarizations.

\begin{figure}[htbp]
\begin{center}
\includegraphics[width=20.5pc]{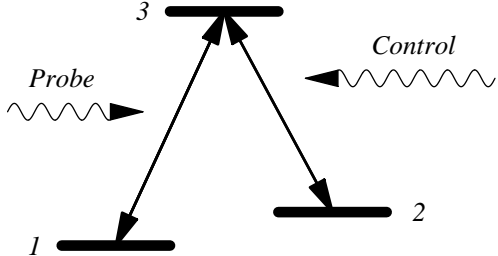}
\vspace*{2mm}

\caption{Physics behind Electromagnetically-Induced
Transparency$^{13}$ (EIT). The figure shows the relevant energy
levels of each atom constituting the EIT medium. The control
light couples two excited states $|\,2\,\rangle$ and
$|\,3\,\rangle$ and thus influences the optical transition
between the ground state $|\,1\,\rangle$ and the level
$|\,3\,\rangle$ brought about by the probe light. Initially, the
control light, being sufficiently strong, prepares each atom in a
pure state $|\,\psi\,\rangle$ called a dark state$^{13,14}$. When
the control and probe field strengths vary the atoms remain in
dark states as long as level $|\,3\,\rangle$ is not sufficiently
populated. Up to a normalization and phase factor, the dark state
$|\,\psi\,\rangle$ is proportional to $|\,1\,\rangle -
(\Omega_p/\Omega_c)\,|\,2\,\rangle + 2(N_0^2/\Omega_c^*)\,
i\partial(\Omega_p/\Omega_c)/\partial t\, |\,3\,\rangle$ with
$N_0^{-2} = 1 + |\,\Omega_p/\Omega_c\,|^2$, described here in an
interaction picture with respect to the atomic transition
frequencies $\omega_{32}$ and $\omega_{31}=\omega_0$. The field
strengths of the probe and control light are given in terms of the
local Rabi frequencies$^{22}$ $\Omega_p$ and $\Omega_c$. The
induced dipole moments of the atoms in dark states generate a
matter polarization that influences the propagation of the probe
light. When $|\,\Omega_p\,|^2$ is much less than
$|\,\Omega_c\,|^2$ the probe light obeys the linear wave equation
(\ref{wave}) with a group index$^{15}$ $\alpha = c/v_g - 1$ that
is inversely proportional$^{14}$ to $|\,\Omega_c\,|^2$. The less
intense the control field is the lower is the group velocity
$v_g$. When $|\,\Omega_p\,|^2$ is comparable with
$|\,\Omega_c\,|^2$ or larger non-linear optical effects occur.}
\label{figure2}
\end{center}
\end{figure}

To predict the propagation properties of slow light we may use
the specific physics of EIT illustrated in figure 2. Equivalently
and more generally, we translate the phenomenological dispersion
relation$^{15}$ of slow light with frequencies close to the
EIT-resonance $\omega_0$ into a wave equation that is subject to
the principle of least action of the canonical formalism$^{16}$,
%%%%%%
\begin{equation}
\label{wave} \left(\frac{\partial}{\partial t}\,(1 +
\alpha)\,\frac{\partial}{\partial t} - c^2
\frac{\partial^2}{\partial z^2} + \alpha\,\omega_0^2 \right)
\varphi = 0 \,.
\end{equation}
%%%%%%
After the formation of the interface ${\cal Z}$ the group index
$\alpha$ has developed a quadratic singularity
%%%%%%
\begin{equation}
\label{alpha} \alpha = c/v_g - 1 = \frac{a^2}{z^2} \,.
\end{equation}
%%%%%%
As a consequence, slow light propagates independently on the two
sides of the interface ${\cal Z}$ and can never cross ${\cal Z}$,
because, in mathematical terms, we can multiply any solution
$\varphi$ with the step function $\Theta(\pm z)$ and still solve
the wave equation (\ref{wave}). On either side of ${\cal Z}$ we
can decompose a slow-light pulse into monochromatic waves, {\it
i.e.} into stationary solutions of the wave equation (\ref{wave})
%%%%%%
\begin{equation}
\label{phi} \varphi = \sqrt{z}\,J_{\pm\nu}(kz)\,e^{\displaystyle
-i\omega t} \,,\quad \nu  =  \sqrt{1/4-a^2(k^2-k_0^2)}
\end{equation}
%%%%%%
expressed in terms of the Bessel functions$^{17}$ $J_\nu$ and the
wave numbers $k=\omega/c$. Two cases emerge. First, when
$4a^2(\omega^2-\omega_0^2) \le c^2$, the index $\nu$ is real. In
this case, the incident waves are totally reflected away from the
interface ${\cal Z}$, as we infer from the behavior of the Bessel
functions$^{17}$ for large $kz$. In the other case,
$4a^2(\omega^2-\omega_0^2)
> c^2$, the index is imaginary,
%%%%%%
\begin{equation}
\nu = i\mu \,,\quad \mu  =  \sqrt{a^2(k^2-k_0^2)-1/4} \,,
\label{mu}
\end{equation}
%%%%%%
and the reflected and incident waves are not balanced. The
remaining transmitted light is trapped at the interface ${\cal
Z}$, because here$^{17}$
%%%%%%
\begin{equation}
\varphi \propto \zeta^{i\mu+1/2}\,e^{-i\omega t} =
\sqrt{\zeta}\,e^{i\mu\ln\zeta-i\omega t} \,,\quad \zeta =  kz \,.
\label{zeta}
\end{equation}
%%%%%%
Close to ${\cal Z}$ the slow-light intensity falls with falling
distance $z$ and the phase $\mu\ln(kz)$ becomes infinite, causing
the light to oscillate with linearly decreasing wave length$^{7}$
$(2\pi/\mu)z$. Waves freeze near the interface ${\cal Z}$. We
regard a process that creates an interface where waves separate
and develop a logarithmic phase singularity$^{18}$ as a {\it wave
catastrophe}.

Close to the event horizon of a black hole$^{2}$ an outside
observer would see a similar behavior of waves$^{19}$. The horizon
cuts space into two disconnected parts. All motion freezes near
the horizon where time seems to stand still. Waves develop a
logarithmic phase singularity. Yet an observer falling inwards
could pass the horizon without noticing anything unusual. This
characteristic difference in perception has a profound
consequence, because the quantum vacuum behaves similar to a
fluid that shares the fate of the inward-falling
observer$^{19-21}$. Consequently, the quantum vacuum must not
occupy the waves seen by the outside observer. In other words,
this observer does not see a vacuum. Instead, the observer
detects the quanta of Hawking radiation$^{3,19}$.

Consider the quantum physics of our wave catastrophe. According to
quantum field theory$^{16}$, waves are potential particle carriers
called modes. Modes describe the spatial-temporal fields of
single quanta and, therefore, they are normalized with respect to
a characteristic scalar product$^{20}$. We normalize the waves
(\ref{phi}) with imaginary index (\ref{mu}) to find the set of
modes
%%%%%%
\begin{eqnarray}
w_R & = & \frac{1}{\sqrt{1-e^{-2\pi\mu}}} \,\left( u_R^+ -
e^{-\pi\mu}\,u_R^-\right) \,,\,\, w_L(z) = w_R(-z) \,,\nonumber\\
u_R & = & u_R^- \,,\,\, u_L(z) = u_R(-z) \,,\nonumber\\
u_R^\pm & = &\frac{\Theta(z)}{\sqrt{2c}}\,e^{-\mu\pi/2}\,
\sqrt{k_0 z}\,J_{\pm i\mu}(kz)\,e^{-i\omega t} \,. \label{set}
\end{eqnarray}
%%%%%%
The step function $\Theta$ indicates that the $R/L$ modes exist
either on the right or on the left side of the horizon ${\cal Z}$.
We have chosen the $w$ modes such that they appear as outgoing
plane waves far away from ${\cal Z}$. These are the modes that
carry detectable quanta.

Complex analysis$^{18}$ is an imaginative mathematical method to
understand real physics. Regard hypothetically the distance $z$
and the time $t$ as complex variables$^{18}$. The modes
(\ref{set}) are non-analytic functions$^{18}$ of $z$, because
they vanish$^{18}$ on one of the sides of the horizon. However,
the quantum vacuum should occupy a set of analytic modes, for the
following reason: The formation of the horizon is a dynamic
process$^{14}$. Initially, the quantum vacuum occupies analytic
modes such as packets of plane waves. The process conserves the
analyticity in $z$, even after the wave catastrophe has occurred.
Furthermore, the light waves are analytic on the lower half of
the complex $t$ plane on at least one side of the horizon. One
can prove that at any arbitrary time $t_0$ after the catastrophe
the vacuum modes are certain combinations of the detector modes,
%%%%%%
\begin{eqnarray}
v_R & = & \frac{w_L}{e^{\pi\mu} + e^{-\pi\mu}} - iw_R
-ie^{-2i\omega t_0} \,\frac{w_R^*}{e^{\pi\mu} + e^{-\pi\mu}}
\,,\nonumber\\
v_R^\perp & = & \frac{1}{\sqrt{1 + e^{2\pi\mu}}}\,\left(u_R +
ie^{\pi\mu} u_L\right) \,, \label{vacuum}
\end{eqnarray}
%%%%%%
supplemented with analogous formulas where $R$ and $L$ are
interchanged. The vacuum modes (\ref{vacuum}) contain
complex-conjugated $w$ modes with negative frequencies and hence
negative energies. This is the decisive sign of particle
creation$^{19,20}$.

\vspace*{1mm}
\begin{figure}[htbp]
\begin{center}
\includegraphics[width=20.5pc]{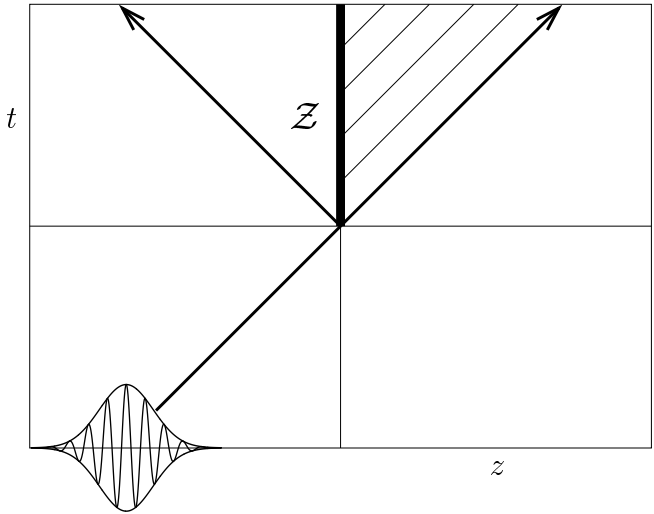}
\vspace*{2mm} \caption{Space-time diagram of a slow-light
catastrophe. The figure illustrates the fate of a wave packet
$\varphi(t,z)$ that experiences the formation of the horizon
${\cal Z}$. Initially, the packet oscillates with positive
frequencies in time $t$ and propagates from the left to the right
in space $z$. The horizon cannot generate negative frequencies in
the reflected light, apart from a brief burst that we neglect. On
the left side of ${\cal Z}$ we thus regard $\varphi(t,z)$ as
analytic$^{18}$ in $t$ on the lower half of the complex $t$ plane.
Furthermore, $\varphi(t,z)$ is analytic in $z$ on the upper half
plane throughout the history of the wave packet, because the
process (\ref{wave}) conserves analyticity$^{18}$. Yet
$\varphi(t,z)$ is not analytic in $t$ on the other side of the
horizon, as the solution (\ref{vacuum}) indicates. Here waves
with negative frequencies are continuously peeling away from the
horizon, corresponding to a stationary creation of slow-light
quanta.} \label{figure3}
\end{center}
\end{figure}

Similar to a gravitational collapse$^{2}$, the tuning of the
control field towards a parabolic intensity profile triggers a
wave catastrophe. In turn, the slow-light quantum field sets out
to deplete the control beam, taking energy from it, in an attempt
to alter the intensity profile that has caused the catastrophe,
yet in vain. The control beam continuously replenishes the
profile, driving a stationary production of slow-light quantum
pairs. The two particles of each pair are created on opposite
sides of the horizon, they depart at a snail's pace, accelerate
gradually and emerge as detectable photons, similar to the
Hawking radiation$^{3,19}$ of black holes. In contrast to
gravitational holes, one can explore the other side of the
horizon and measure the non-local correlations$^{22}$ of the
photon pairs. The weight of the negative-frequency component
$w^*$ in the vacuum modes (\ref{vacuum}) gives$^{20}$ the average
photon number per mode,
%%%%%%
\begin{equation}
\label{leo} \bar{n} = \frac{1}{\left(e^{\pi\mu} +
e^{-\pi\mu}\right)^2} \,.
\end{equation}
%%%%%%
Maximally $1/4$ photons are created on average, which is quite
substantial, considering the fact that bright sunlight carries a
mere $0.01$ photons per mode in the optical range of the Planck
spectrum$^{22}$. Yet the pair production occurs in a narrow
frequency window above the critical frequency $\omega_0 +
c^2/(8a^2\omega_0)$ which, for realistic experimental
parameters$^{4,6}$, reduces the total photon flux to a few
millions of particles per second, a respectable rate. One could
perhaps see the radiation with the naked eye. The characteristic
length $a$ of the group-velocity profile (\ref{v}) determines the
spectral width (\ref{leo}) of the pair production. The steeper the
control-field gradient is, the more quanta are created. Black
holes show a similar behavior$^{3}$. The smaller the hole is, the
larger is the gravity gradient at the horizon and the stronger is
the radiation generated$^{3}$.

Close to the horizon the susceptibility of slow light diverges.
Yet Nature tends to prevent infinite susceptibilities: Instead of
responding infinitely strongly, optical media become absorptive or
non-linear. According to the physics of EIT illustrated in figure
2, the optical non-linearity of slow light depends on the ratio
of the probe and control intensities $I_p$ and $I_c$. Equation
(\ref{zeta}) shows that $I_p \propto |\,\varphi\,|^2$ grows
linearly in $z$, whereas $I_c$ is quadratic in $z$. Therefore, at
a critical distance from the horizon the EIT medium becomes
non-linear. A detailed three-dimensional calculation, using the
parameters of the experiments$^{4,6}$, indicates that the
non-linearity sets on before the absorption of the medium becomes
important, given a sufficiently steep control-intensity gradient.
The Rabi frequency$^{22}$ of the control light should grow at
least by $10 {\rm MHz}$ per distance measured in wave lengths
$\lambda_0=2\pi/k_0$. In this case the scale $a$ is about
$5\times10^3\lambda_0$.

The quantum radiation of a slow-light catastrophe resembles
Hawking radiation but also exhibits some interesting differences.
The emitted spectrum (\ref{leo}) is not Planckian, whereas a black
hole of Schwarzschild radius $r_s$ appears as a black-body
radiator with temperature$^{3}$ $\hbar c/(4\pi r_s)$. The
differences between the two spectra can be traced back to two
different classes of wave catastrophes. In both cases$^{18}$,
waves freeze at an horizon in the form $\zeta^p$ with an exponent
$p$ of $i\mu + {1}/{2}$ for slow-light media but with an exponent
$i\mu$ for black holes where $\mu = 2\pi r_s\, \omega/c$. Note
that Unruh's effect$^{21}$ of radiation seen by an accelerated
observer is of Hawking-class as well$^{19}$ and so are most of the
proposed artificial black holes$^{15,23-29}$. Remarkably,
Schwinger's pair production of charged particles in electrostatic
fields$^{30}$ is accompanied by a subtle wave catastrophe of
exponent$^{19}$ $i\mu - {1}/{2}$ and leads to a Boltzmannian
spectrum $\bar{n}=\exp(-2\pi\mu)$. All three catastrophes agree
in the limit of large $\mu$ but deviate significantly in the
regime of maximal particle production where $\mu$ is small. It
might be interesting to find out whether Nature offers more than
the three quantum catastrophes.

Details of the calculations will be published elsewhere.

%%%%%%%%%%%%%%%%%%%%%%%%%%%%%%%%%%%%%%%%%%%%%%%%%%%%%%%%%%%%%%%%%%%%%%%%%%

%\newpage
\begin{enumerate}

\item
Berry, M. V. \& Upstill, C. Catastrophe optics: morphologies of
caustics and their diffraction patterns. {\it Progress in Optics
XVIII} 257-346 (1980).

\item
Misner, Ch. W., Thorne, K. S. \& Wheeler, J. A. {\it Gravitation}
(Freeman, New York, 1999).

\item
Hawking, S. M. Black hole explosions? {\it Nature} {\bf 248},
30-31 (1974).

\item
Liu, Ch., Dutton, Z., Behroozi, C. H. \& Hau, L. V. Observation of
coherent optical information storage in an atomic medium using
halted light pulses. {\it Nature} {\bf 409}, 490-493 (2001).

\item
Philips, D. F., Fleischhauer, A., Mair, A., Walsworth, R. L. \&
Lukin, M. D. Storage of Light in Atomic Vapor. {\it Phys. Rev.
Lett.} {\bf 86}, 783-786 (2001).

\item
Dutton, Z., Budde, M., Slowe, C. \& Hau L. V. Observation of
quantum shock waves created with ultra-compressed slow light
pulses in a Bose-Einstein condensate. {\it Science} {\bf 293},
663-668 (2001).

\item
Born, M. \& Wolf, E. {\it Principles of Optics} (Cambridge
University Press, Cambridge, 1999).

\item
Thom, R. {\it Stabilit\'{e} structurelle et morphog\'{e}n\`{e}se}
(Benjamin, Reading, 1972).

\item
Poston, T. \& Stewart, I. {\it Catastrophe Theory and Its
Applications} (Dover, Mineola, 1996).

\item
Schleich, W. \& Scully, M. O. General relativity and modern
optics. {\it Les Houches Session XXXVIII New trends in atomic
physics} (Elsevier, Amsterdam, 1984).

\item
Wang, L. J., Kuzmich, A. \& Dogariu, A. Gain-assisted
superluminal light propagation. {\it Nature} {\bf 406}, 277-279
(2000).

\item
Hau, L. V., Harris, S. E., Dutton, Z. \& Behroozi, C. H. Light
speed reduction to 17 metres per second in an ultracold atomic
gas. {\it Nature} {\bf 397}, 594-598 (1999).

\item
Scully, M. O. \& Zubairy, M. S. {\it Quantum Optics} (Cambridge
University Press, Cambridge, 1997).

\item
Fleischhauer, M. \& Lukin, M. D. Dark-state polaritons in
electromagnetically-induced transparency. {\it Phys. Rev. Lett.}
{\bf 84}, 5094-5097 (2000).

\item
Leonhardt, U. \& Piwnicki, P. Relativistic effects of light in
moving media with extremely low group velocity. {\it Phys. Rev.
Lett.} {\bf 84}, 822-825 (2000).

\item
Weinberg, S. {\it The Quantum Theory of Fields} (Cambridge
University Press, Cambridge, 1999).

\item
Erd\'{e}lyi, A., Magnus, W., Oberhettinger, F. \& Tricomi, F. G.
{\it Higher Transcendental Functions}, (McGraw-Hill, New York,
1981).

\item
Ablowitz, M. J. \& Fokas, A. S. {\it Complex Variables} (Cambridge
University Press, Cambridge, 1997).

\item
Brout, R., Massar, S., Parentani, R. \& Spindel, Ph. A primer for
black hole quantum physics. {\it Phys. Rep.} {\bf 260}, 329-446
(1995).

\item
Birrell, N. D. \& Davies, P. C. W. {\it Quantum Fields in Curved
Space} (Cambridge University Press, Cambridge, 1982).

\item
Unruh, W. G. Notes on black-hole evaporation. {\it Phys. Rev. D}
{\bf 14}, 870-892 (1976).

\item
Mandel, L. \& Wolf, E. {\it Optical Coherence and Quantum
Optics}, (Cambridge University Press, Cambridge, 1995).

\item
Unruh, W. G. Experimental black-hole evaporation? {\it Phys. Rev.
Lett.} {\bf 46}, 1351-1353 (1981).

\item
Visser, M. Acoustic black holes: horizons, ergosheres and Hawking
radiation. {\it Class. Quantum. Grav.} {\bf 15}, 1767-1791 (1998).

\item
Jacobson, T.A \& Volovik, G. E. Event horizons and ergoregions in
$^3$He. {\it Phys. Rev. D} {\bf 58}, 064021-1-7 (1998).

\item
Volovik, G. E. Simulation of a Panlev\'e-Gullstrand black hole in
a thin $^3$He-A film. {\it JETP Lett.} {\bf 69}, 705-713 (1999).

\item
Garay, L. J., Anglin, J. R., Cirac, J. I. \& Zoller, P. Sonic
analog of gravitational black holes in Bose-Einstein condensates.
{\it Phys. Rev. Lett.} {\bf 85}, 4643-4647 (2000).

\item
Reznik, B. Origin of the thermal radiation in a solid-state
analogue of a black hole. {\it Phys. Rev. D} {\bf 62}, 044044-1-7
(2000).

\item
Chapline, G., Hohfeld, E., Laughlin, R. B. \& Santiago, D. I.
Quantum phase transitions and the breakdown of classical general
relativity. {\it Phil. Mag. B} {\bf 81}, 235-254 (2001).

\item
Schwinger, J. On gauge invariance and vacuum polarization. {\it
Phys. Rev.} {\bf 82}, 664-679 (1951).

\end{enumerate}

%\newpage
%\widetext

\vspace*{10mm} \noindent
\begin{tabular}{|l|c|c|}\hline
& & \\
\hspace*{7mm} Example & \hspace*{2mm} Exponent \hspace*{2mm} &
Average particle number\\ & & \\
\hline\hline & & \\
\hspace*{7mm} Hawking radiation & $i\mu$ &
\begin{minipage}[b]{20mm}$$\frac{1}{e^{2\pi\mu} - 1}$$
\end{minipage} \\
\hspace*{7mm} Unruh effect & & \\ & & \\\hline  & & \\
\hspace*{7mm} Schwinger's pair production \hspace*{7mm} & $i\mu -
1/2$ & \begin{minipage}[b]{20mm}$$e^{-2\pi\mu}$$
\end{minipage} \\ & & \\
\hline & & \\ \hspace*{7mm} Slow light & $i\mu + 1/2$
&\begin{minipage}[b]{20mm}$$\frac{1}{\left(e^{\pi\mu} +
e^{-\pi\mu}\right)^2}$$\end{minipage} \\
& & \\ \hline
\end{tabular}

\vspace*{10mm} \noindent {\it Quantum catastrophes. In each
example a wave develops a singularity with a characteristic
exponent. Quantum physics resolves the singularity and produces
particle pairs with a characteristic spectrum (average particle
number).}

\newpage

\noindent ACKNOWLEDGEMENTS. The Discussion Meeting on Artificial
Black Holes at the Royal Institution has been an inspiration for
this work. I take this opportunity to thank the participants and
the organizers. I acknowledge the support of the ESF Programme
Cosmology in the Laboratory.\\

\noindent CORRESPONDENCE should be addressed to U.\ L. (e-mail:
ulf@st-andrews.ac.uk).
\end{document}